\pdfoutput=1

\documentclass[11pt]{article}

\usepackage{EMNLP2023}

\usepackage{times}
\usepackage{latexsym}
\usepackage{graphicx}

\usepackage[T1]{fontenc}
\usepackage{amsmath}

\usepackage[utf8]{inputenc}

\usepackage{microtype}

\usepackage{inconsolata}
\usepackage{booktabs}
\usepackage{array}
\usepackage{multirow}
\usepackage{listings}
\usepackage{xcolor}

\usepackage{enumitem}
\usepackage{xspace}

\lstdefinestyle{codestyle}{
    backgroundcolor=\color{gray!10},   
    basicstyle=\ttfamily\small,
    breakatwhitespace=false,         
    breaklines=true,                 
    keepspaces=true,                 
    numbersep=5pt,                  
    showspaces=false,                
    showstringspaces=false,
    showtabs=false,                  
    tabsize=4,
    frame=single,
    framerule=1.5pt,
    xleftmargin=0.5cm,
    xrightmargin=0.5cm,
    aboveskip=1cm,
    belowskip=0.5cm,
    escapeinside={<@}{@>}
}

\newcommand{\hlred}[1]{%
    \setlength{\fboxsep}{0pt}
    \colorbox{red!30}{#1}%
}

\newcommand{\agentname}{\textsc{SimAgents}\xspace}
\def\agentone{Physics Agent\xspace}
\def\agenttwo{Software Agent\xspace}

\title{\agentname: Bridging Literature and the Universe Via A Multi-Agent \\ Large Language Model System}

\author{ \bf
  Xiaowen Zhang\textsuperscript{\(\spadesuit\)}$^*$, 
  Zhenyu Bi\textsuperscript{\(\heartsuit\)}$^*$, 
  Patrick Lachance\textsuperscript{\(\spadesuit\)}, \\
  \bf
  Xuan Wang\textsuperscript{\(\heartsuit\)}, 
  Tiziana Di Matteo\textsuperscript{\(\spadesuit\)}, 
  Rupert A.~C. Croft\textsuperscript{\(\spadesuit\)}\textsuperscript{\textdagger} \\
  \textsuperscript{\(\spadesuit\)}Department of Physics, Carnegie Mellon University, Pittsburgh, PA, USA \\
  \textsuperscript{\(\heartsuit\)}Department of Computer Science, Virginia Tech, Blacksburg, VA, USA \\
 \textsuperscript{\(\spadesuit\)}\texttt{(xiaowen4,plachanc,tizianad)@andrew.cmu.edu}, \texttt{rcroft@cmu.edu}, \\
 \textsuperscript{\(\heartsuit\)}\texttt{(zhenyub,xuanw)@vt.edu}
}

\begin{document}
\maketitle
\def\thefootnote{*}\footnotetext{Equal contribution}\def\thefootnote{\arabic{footnote}}
\def\thefootnote{\textdagger}\footnotetext{Corresponding author.}\def\thefootnote{\arabic{footnote}}
\begin{abstract}
As cosmological simulations and their associated software become increasingly complex, physicists face the challenge of searching through vast amounts of literature and user manuals to extract simulation parameters from dense academic papers, each using different models and formats. Translating these parameters into executable scripts remains a time-consuming and error-prone process. To improve efficiency in physics research and accelerate the cosmological simulation process, we introduce \agentname, a multi-agent system designed to automate both parameter configuration from the literature and preliminary analysis for cosmology research. $\agentname$ is powered by specialized LLM agents capable of physics reasoning, simulation software validation, and tool execution. These agents collaborate through structured communication, ensuring that extracted parameters are physically meaningful, internally consistent, and software-compliant. We also construct a cosmological parameter extraction evaluation dataset by collecting over 40 simulations in published papers from Arxiv and leading journals that cover diverse simulation types. Experiments on the dataset demonstrate a strong performance of \agentname, highlighting its effectiveness and potential to accelerate scientific research for physicists. Our demonstration video is available at: https://youtu.be/w1zLpm\_CaWA. The complete system and dataset are publicly available at https://github.com/xwzhang98/SimAgents.
\end{abstract}

\section{Introduction}


Modern cosmological simulations are essential tools for advancing our understanding of the universe, enabling researchers to study the formation of galaxies and the evolution of structures. Setting up such simulations is a highly manual, time-consuming, and error-prone process. Researchers must extract parameters from dense scientific papers, convert values between units, interpret context-specific model assumptions, and then format them into executable scripts compatible with domain-specific software such as MP-GADGET \citep{mp-gadget}, GADGET-4 \citep{gadget4}, Arepo \citep{arepo}, GIZMO code \citep{gizmo} and ENZO \citep{enzo}. 
In addition to the diversity of the simulations themselves, the complexity of using the software adds another layer of difficulty. Software user manuals are often dozens of pages long and filled with intricate rules about parameter dependencies, default settings, and strict formatting requirements.

\begin{figure*}[!t]
\centering
\includegraphics[scale=0.70]{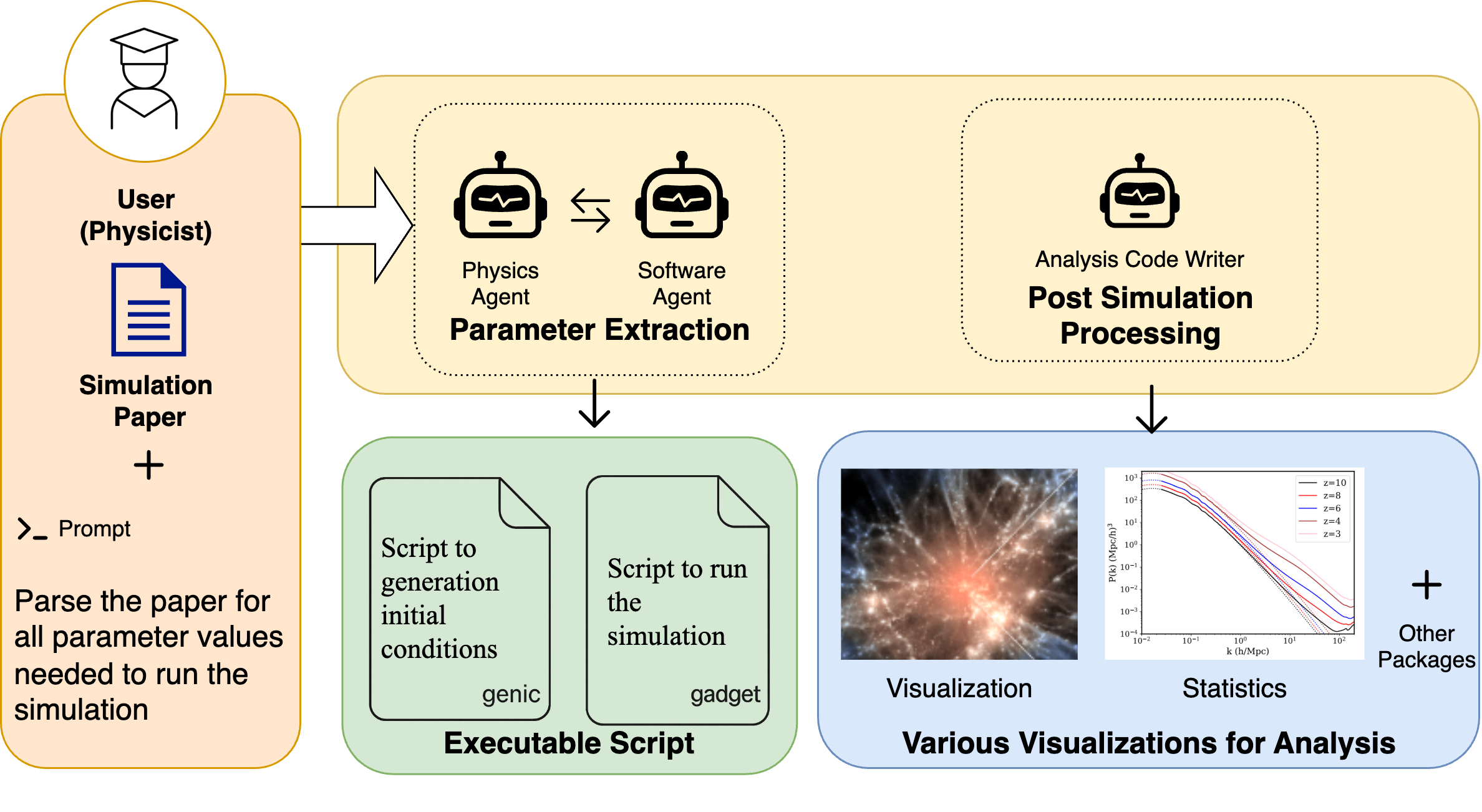}
\caption{The workflow of our proposed multi-agent system, \agentname.}
\label{fig:Structure}
\end{figure*}

As a result, even experienced physics researchers face a steep learning curve when trying to adopt a new simulation tool. For example, when given a cosmology paper covering several simulations, the average time cost for a human researcher to formulate the correct parameter files is in the range of hours to days, depending on the familiarity with the software. Ideally, we want this labor-intensive process to be done in minutes.
The above challenges raise a crucial question: \textbf{How can we design a highly professional automated toolkit to assist cosmologists with the lengthy and complex task of setting up simulations?} 

Large Language Model (LLM) agents have demonstrated significant potential on many scientific tasks \citep{LLMsurvey}. Recently, researchers have proposed multi-agent reasoning frameworks that enable collaborative debates among multiple LLM agents to enhance their problem-solving abilities \citep{AutoGen,MAD,GPTSwarm,stoctot}. Following this path, researchers have explored LLM-agent-based workflows on several highly professional scientific and technical applications, such as biomedical tasks and clinical tasks \citep{biomedsurvey,triargeagent}.

In the field of cosmology, researchers have explored various LLM agent tools to provide assistance to researchers, targeting several tasks, such as a programming assistant specialized in different cosmology tasks. For example, CLAPP\footnote{\url{https://github.com/santiagocasas/clapp/}} is a single LLM agent that specializes in the CLASS cosmology code. CAMEL agents \footnote{\url{https://github.com/franciscovillaescusa/CAMELS_Agents/}} provide a suite of AI-powered agents designed specifically to navigate and analyze the extensive CAMELS cosmological simulation dataset, automating tasks such as data exploration and code generation. In addition, CMBAgent \citep{cmbagent} and Mephisto \citep{multi-band-agent} utilize a multi-agent LLM system to aid physicists in cosmological parameter analysis. Each of these systems focuses on a different scope of research, ranging from coding support to data analysis research directions. \textbf{However, to our knowledge, no prior LLM agent system automates the whole workflow from parameter configuration from the literature to initial simulation output analysis on cosmology simulation software.}

Toward this end, we introduce \agentname, a multi-agent system that automates parameter extraction, validation and configuration for cosmological simulations. The system is composed of specialized LLM agents with different distinct roles:
\begin{itemize}[leftmargin=*] 
    \setlength\itemsep{-0.9em}
    \setlength\parskip{1em} 
    \item \textbf{\agentone} that reads and interprets simulation papers using domain knowledge
    \item \textbf{\agenttwo} that parses and enforces the constraints specified in the software user manual
    \item \textbf{Analysis Code Writer} that provides codes for result visualization and produces preliminary analysis (e.g. power spectra and density fields plot)
\end{itemize}
These agents collaborate through structured communication, ensuring that extracted parameters are physically meaningful, internally consistent, and software-complaint. 
To assess the effectiveness of \agentname, we construct a benchmark dataset of 41 simulations and evaluate the system's performance using metrics such as precision and recall, and error-specific breakdowns (e.g. Value Error, Type Error and Hallucinations). Our results show that \agentname achieves high accuracy while significantly reducing the manual workload typically required for simulation setup.

\section{\agentname}
In this section, we present the structure and implementations of \agentname. As illustrated in Figure \ref{fig:Structure}, $\agentname$ is composed of the following key components:

\begin{itemize}[leftmargin=*]
    \setlength\itemsep{-0.9em}
    \setlength\parskip{1em} 
    \item \textbf{Parameter extraction:} This module automates the process of generating simulation scripts by extracting relevant parameters from user-uploaded papers and formatting them according to the internal requirements of the target simulation software. The extraction is performed through iterative communication between a dual-agent setup, ensuring accuracy and consistency.
    \item \textbf{Post Simulation Processing:} This module handles code generation and execution for preliminary simulation analysis, including power spectra and density field plotting.
\end{itemize}

Together, the simulation preparation and preliminary analysis step allows users to move quickly from a published paper to actionable simulation output, closing the loop from literature reading to research insight.

\subsection{Parameter Extraction}
The parameter extraction module is responsible for transforming scientific papers into structured simulation-ready configurations. Given a user-uploaded paper, the system initiates a dual-agent collaboration between \textbf{\agentone} and \textbf{\agenttwo}.
The \textbf{\agentone} reads the input paper using domain knowledge in cosmology to identify relevant parameters such as cosmological constants, simulation box size, redshift and simulation types (dark matter, gas, stars and neutrinos). The \textbf{\agenttwo} utilizes the simulation software's official user manual to query all required and optional parameters, including their default values, units, and inter-parameter dependencies.

These two agents collaborate through participation in multiple rounds of discussions based on the provided material, including a research paper and the software manual, to refine the parameter extraction process. Specifically, after \agentone reads the paper and extracts the parameters, the results will be sent to the \agenttwo, which will then use the software user manual to check the coverage and validity of the extracted parameters. Then \agenttwo will generate the parameter file following the required format and constraints. The generated file will then be sent to \agentone for another round of refinement. This iterative process, together with specialized task assignment on each agent, ensures:
\begin{itemize} [leftmargin=*]
    \setlength\itemsep{-0.9em}
    \setlength\parskip{1em} 
    \item \textbf{High accuracy}, including scientific parameter accuracy and software requirement compliance, through task-specific expertise;
    \item \textbf{Modular adaptability}, as the formatting agent can be extended to support different simulation software by referencing alternative user manuals without altering the extraction logic.
\end{itemize}

\subsection{Post-Simulation Processing}

Once parameter extraction is completed, the generated script is passed to the simulation software for execution. After obtaining the output, the system transitions to the post-simulation processing stage, where an \textbf{Analysis Code Writer} automatically generates Python scripts to assist users with early-stage analysis of the simulation output. These generated scripts support:
\begin{itemize} [leftmargin=*] 
    \setlength\itemsep{-0.9em}
    \setlength\parskip{1em}
    \item \textbf{Visualization:} Generating 2D/3D density plots from slices of the simulation box.
    \item \textbf{Statistical Analysis:} Generating code to plot summary statistics like matter power spectrum.
    \item \textbf{Custom Post-Processing:} Capability to use user-provided custom packages
\end{itemize}

The scripts are designed to be executable with minimal modification and make use of standard Python libraries such as NumPy, Matplotlib. For specialized cosmology packages, the system generates code based on example usage provided to the agent. This stage helps researchers validate simulations, identify issues early and prepare for deeper scientific investigation.

\section{Experimental Setup}
Our experiments are conducted in two parts: the first focuses on parameter extraction, where we evaluate quantitative accuracy; the second addresses simulation post-processing, which is more subjective and demonstrated through a representative pipeline. In our paper, we use MP-GADGET as our simulation software. In the following, we describe the experimental setup for parameter extractions.
\label{sec:exp}

\paragraph{Dataset}
We construct a dataset for the evaluation of cosmological parameter extraction by collecting more than 40 different simulations from published articles from ArXiv and leading journals (e.g. \emph{ApJ}, \emph{MNRAS}). To run MP-GADGET, two input files are required: a .genic file and a .gadget file. The .genic file generates the initial positions and velocities of particles, along with essential simulation metadata. The .gadget file evolves the initial particle distribution over time and contains numerous configuration options for selecting and enabling various physical models. Each paper is manually annotated with all MP-GADGET relevant parameter value pairs, covering cosmological parameters ($\Omega_{m}$, $\Omega_{b}$, $\Omega_{\Lambda}$, $h$, $\sigma_{8}$, $n_{s}$), initial‐condition settings (\texttt{BoxSize}, \texttt{Ngrid}, \texttt{Redshift}), and key model switches (e.g.\ \texttt{StarformationOn}, \texttt{WindOn}). 
To our knowledge, this is the first publicly released dataset of cosmological simulations with parameters derived directly from published text.

\paragraph{Implementation}
We use OpenAI GPT-4 \citep{GPT4} for our zero-shot extraction experiments.  Our \texttt{\agentname} framework utilizes the publicly available Autogen framework\footnote{\url{https://microsoft.github.io/autogen/}}. We also conduct an ablation study of our \agentname using the Qwen3-4B model \citep{qwen3}. We set the temperature to 0.01 and $top\_p$ to 0.1. For the simulation software, we use MP-GADGET as an example in this paper. All outputs are formatted directly in MP-GADGET configuration syntax. We conduct all the experiments with user manual since the LLM does not have sufficient knowledge of current simulation software.

\paragraph{Baselines} We compare our methods against two baseline methods.
\begin{itemize}[leftmargin=*]
    \setlength\itemsep{-0.9em}
    \setlength\parskip{1em} 
     \item \textbf{Chain-of-thought (CoT)} \citep{chainofthought} We implement zero-shot CoT prompting with a single LLM agent. The agent is provided with both the literature and the manual.
     
     \item \textbf{Exchange-of-thought (EoT)} \citep{exchangeofthought}  We implement EoT using two agents with the same initialization, and provide them both with the literature and the manual. The agents engage in a discussion with one another.
     
    \item \textbf{\agentname} Our approach employs two task‐specific agents: \agentone and \agenttwo, each with role-specialized profiling. We provide \agentone with only the literature and \agenttwo with only the manual. The agents engage in a discussion with one another.
\end{itemize}

We recognize that there are other LLM-based retrieval augmented generation frameworks \cite{RAGsurvey}. However, these RAG methods are unnecessary for our current work, as the information we provide is straightforward and does not need special design on the RAG techniques. Other LLM-based multi-agent tools in the field of cosmology \citep{cmbagent,multi-band-agent} do not fit into the scope of the current work. Thus, we do not compare with these methods in our baselines.

\paragraph{Evaluation} We evaluate our framework using F1-score and different error metrics and provide the details of these metrics in Appendix \ref{Appendix: Eval}. Due to time constraints, we only annotated one version of the executable files. For each simulation, there exist multiple variants that contain parameters not covered in the original paper, but which could still yield the same output. To facilitate a fair comparison with the baselines, we conduct a human evaluation covering as many variants as possible and report the results in Table \ref{tab:accuracy} and Table \ref{tab:error-analysis}. The automated evaluation against the annotated dataset is reported in Section \ref{Appendix: autoEval}.

\section{Results}
In this section, we first present the quantitative results, which contains baseline comparisons, detailed error analysis, ablation studies, and cost analysis. We then present a brief overview of the post-simulation processing capabilities of our system.

\subsection{Main Results}
\begin{table}[t]\normalsize
\centering
\setlength{\tabcolsep}{6pt} 
\begin{tabular}{@{}lccc@{}}
\toprule
\textbf{Method} & \textbf{Micro-F1} & \textbf{Precision} & \textbf{Recall} \\
\midrule
CoT (1-Agent)  & 93.64 & 92.46 & 94.84 \\
EoT (2-Agent)  & \underline{94.95} & \underline{93.87} & \underline{96.05} \\
Ours (2-Agent) & \textbf{98.67} & \textbf{97.80} & \textbf{99.55} \\
\bottomrule
\end{tabular}
\caption{Performance comparison of \agentname with baseline methods on the cosmological simulation dataset. We report Micro-F1 score, Precision, and Recall as percentages. \textbf{Higher values indicate better performance.} The best-performing methods are bolded, and the second-best are underlined.}
\label{tab:accuracy}
\end{table}

\begin{table}[t]\normalsize
\centering
\setlength{\tabcolsep}{4pt} 
\begin{tabular}{@{}lccc@{}}
\toprule
\textbf{Method} & \textbf{Value} & \textbf{Type} & \textbf{Hallucination} \\
                & \textbf{Error} & \textbf{Error} &  \\
\midrule
CoT (1-Agent)  & \underline{0.97} & 0.51 & \textbf{0.21} \\
EoT (2-Agent)  & 1.21 & \underline{0.21} & 0.34 \\
Ours (2-Agent) & \textbf{0.46} & \textbf{0.02} & \underline{0.30} \\
\bottomrule
\end{tabular}
\caption{Performance comparison of \agentname with baseline methods on the cosmological simulation dataset, in terms of average number of errors made per simulation. Each error type is reported as the average number of errors per case. \textbf{Lower values indicate better performance.} The best-performing methods are bolded, and the second-best are underlined.}
\label{tab:error-analysis}
\end{table}

The performance of our system compared to the baseline methods is shown in Table \ref{tab:accuracy}. Our proposed method, \agentname, outperforms CoT and EoT, achieving improvements of 5.03\% and 3.72\% in Micro-F1 score, respectively. Reduces the overall error rate by 80\% compared to CoT and 70\% compared to EoT, demonstrating significantly improved reliability in parameter extraction.

\paragraph{Comparison with Baseline Methods}
We examine the reasoning process of the CoT method and find that it struggles to handle excessive task instructions and information at input time, consistently making errors, and is unable to complete any tasks effectively. Simply involving multiple agents is not sufficient for optimal performance: EoT benefits from multi-agent interaction, but its lack of specialized task decomposition and clear communication structures results in imprecise outcomes. In contrast, \agentname incorporates task-specialized agents with specialized inputs, significantly reducing critical error types and leading to more accurate and robust parameter extraction.

\begin{figure}[!t]
\centering
\includegraphics[width=\columnwidth]{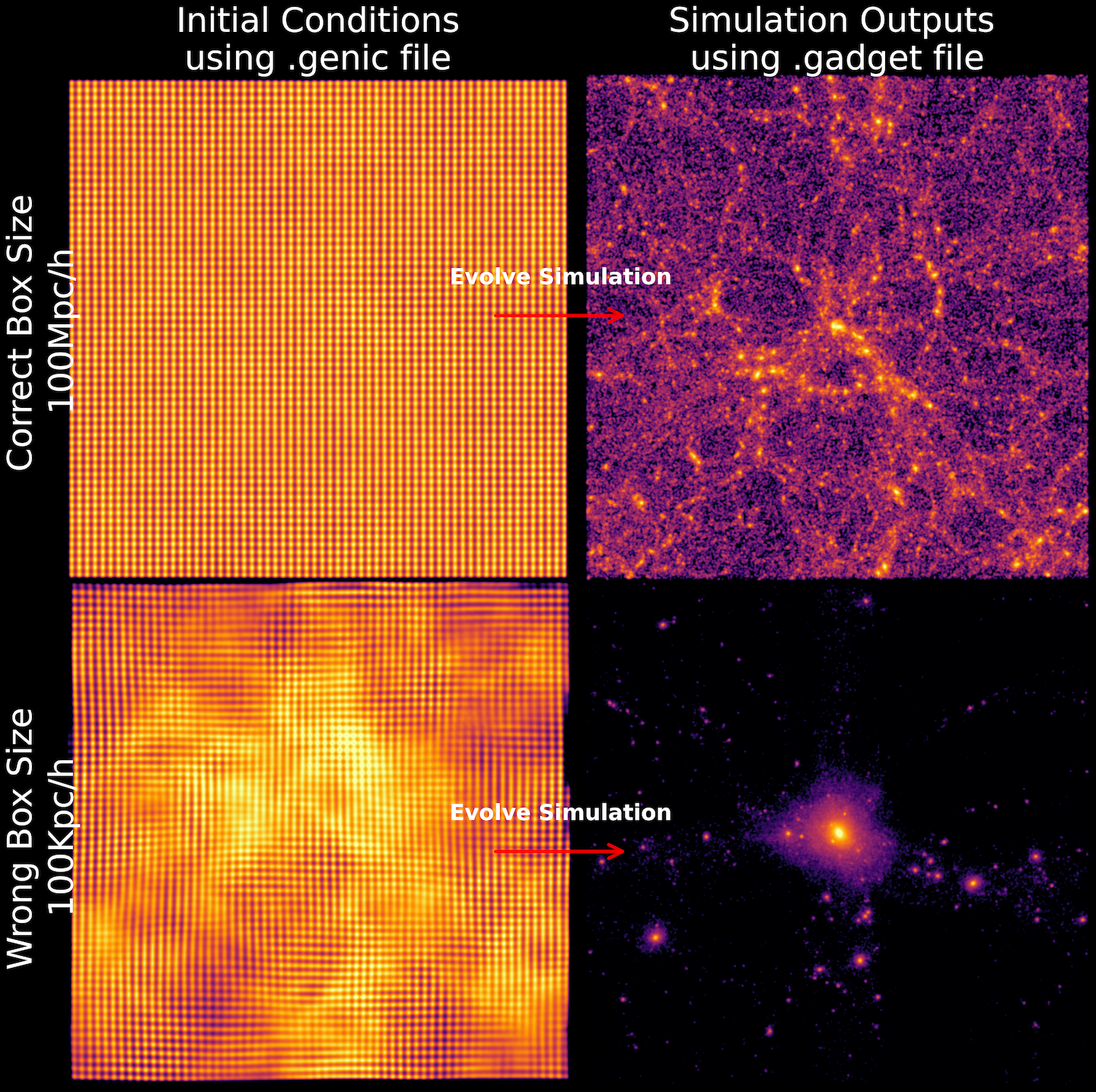}
\caption{Impact of incorrect parameters (Value Error) on cosmological simulation outputs. Varying a single parameter, such as box size (correct top row: 100 Mpc/h; incorrect bottom row: 100 Kpc/h), while keeping all others fixed, can result in drastically different structures.}
\label{fig:parameter_importance}
\end{figure}

\paragraph{Error Analysis}
In detailed error analysis, we observe that both CoT-based extraction and EoT-based extraction exhibits a higher frequency of both value error and type errors as shown in Table \ref{tab:error-analysis}. Although our system exhibits slightly higher hallucination per case than the other baselines, these hallucinated parameters are easier to detect and filter (we provide an example in Appendix \ref{appendix_example_script}). In contrast, value errors involve plausible-looking parameters whose values or units are subtly incorrect, often bypassing sanity checks and undetected during the simulation stage. Figure \ref{fig:parameter_importance} shows that a single value error leads to drastically different structures, due to the different unit convention between the literature and simulation software.

\begin{figure}[!t]
\centering
\includegraphics[width=0.9\columnwidth]{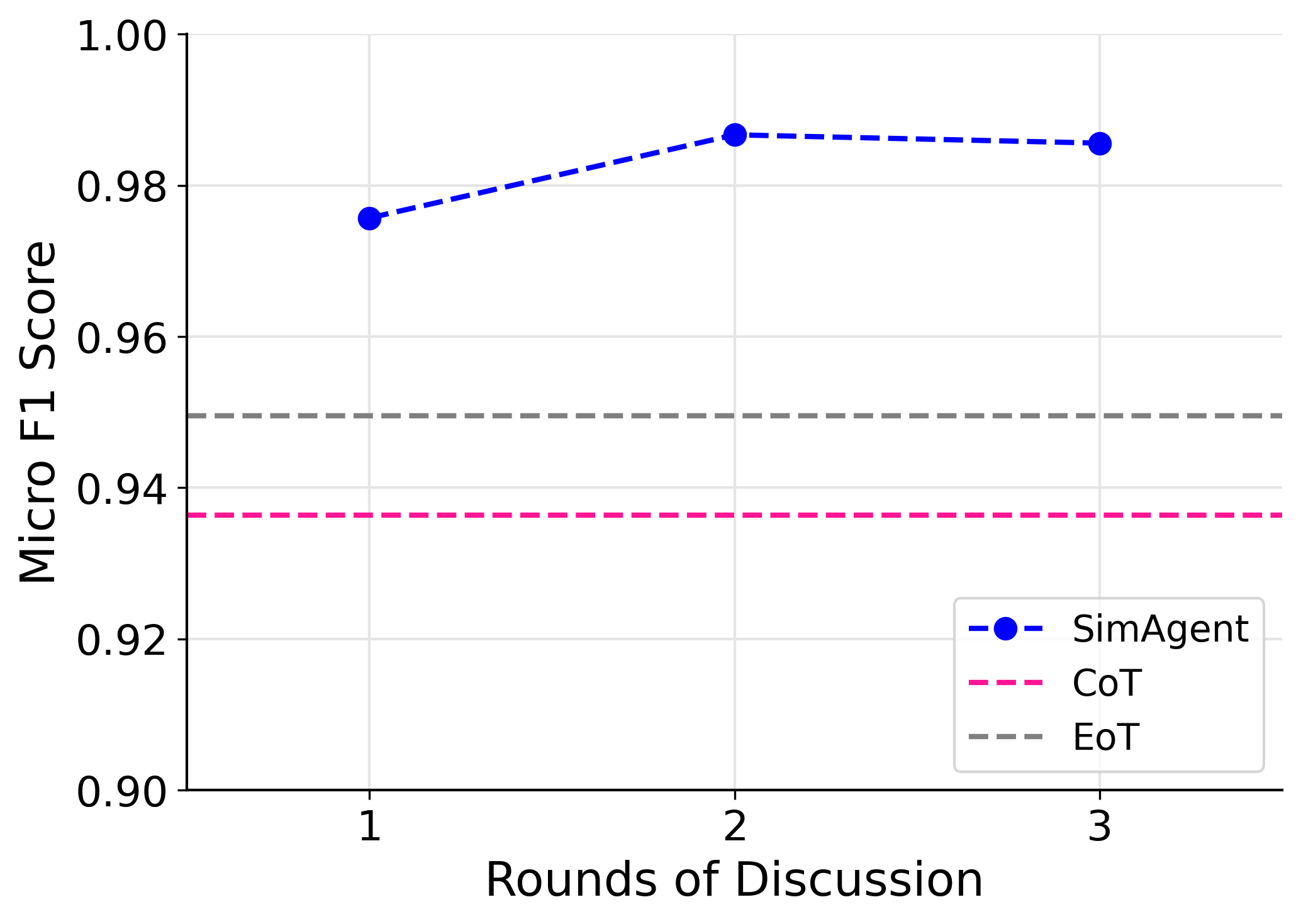}
\caption{Results for the ablation study on the number of rounds of discussion.}
\label{fig:iteration_impact}
\end{figure}

\begin{figure*}[!t]
\centering
\includegraphics[scale=0.85]{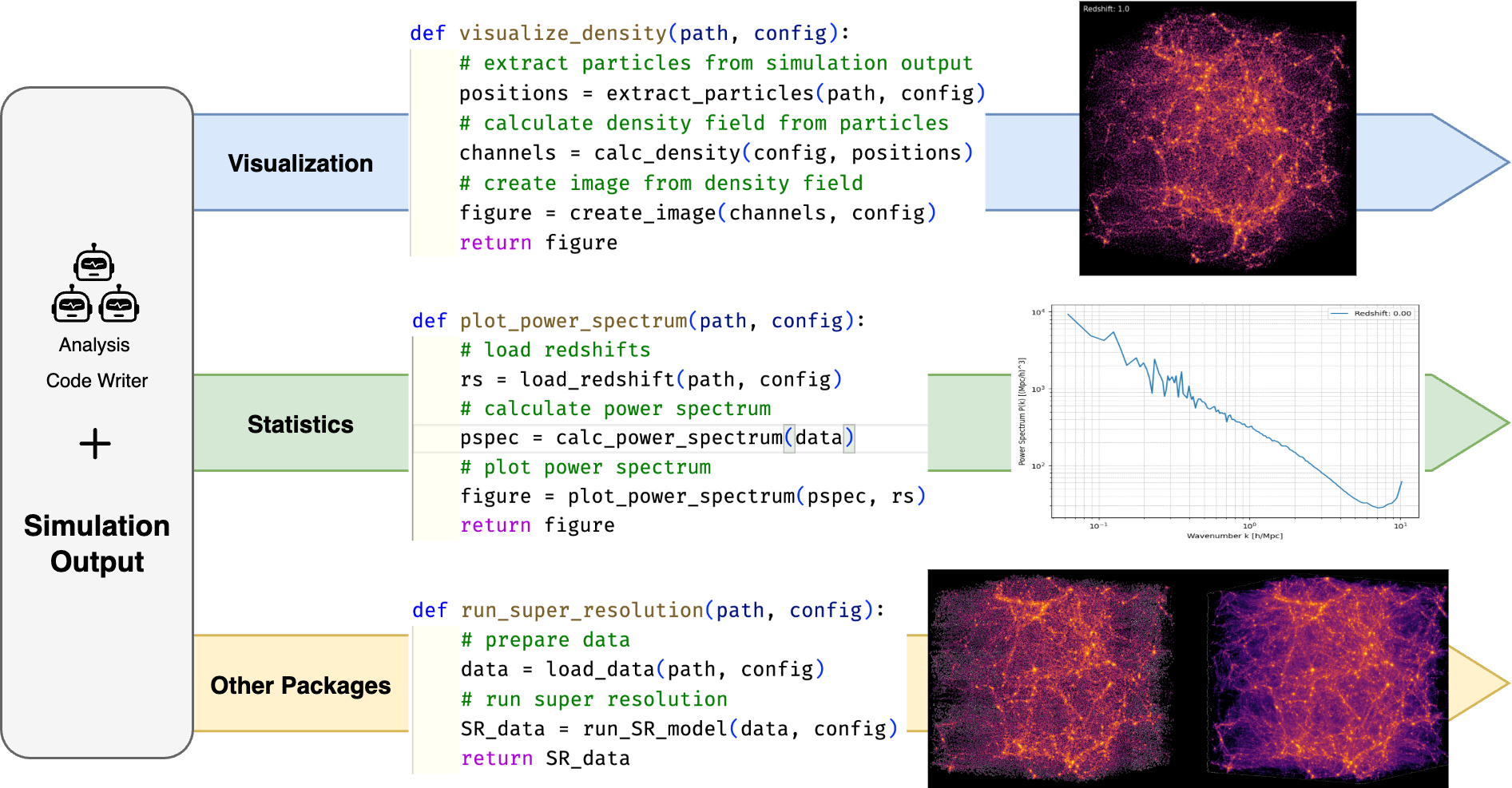}
\caption{Illustration of post-simulation processing pipeline}
\label{fig:post_processing}
\end{figure*}

\subsection{Ablation study}
\paragraph{Rounds of Discussion}
We conduct an ablation study to investigate the optimal number of discussion rounds between \agentone and \agenttwo. In our parameter extraction module, each agent contributes domain-specific expertise to achieve high extraction accuracy while maintaining computational efficiency. By varying the number of discussions between these agents, we observe that two iterations yield the highest Micro-F1 score, as shown in Figure \ref{fig:iteration_impact}. 
\paragraph{Smaller Backbone Model} We also conduct experiments using Qwen3-4B as the backbone model to examine the generalizability of \agentname on Small Language Models. We provide the detailed results in Table \ref{tab:qwen_accuracy} and Table \ref{tab:qwen_error-analysis} in Appendix \ref{Appendix: autoEval}. Compared with GPT-4 which is significantly larger in model size, Qwen3-4B has inferior reasoning ability, leading to a decreased performance of an 81.23 F1 score, and an average of 3.05 value errors and 2.59 type errors per simulation. 

\subsection{Time and Cost Analysis}
Compared to a human researcher who typically spends hours (or even more if unfamiliar with the code) parsing a paper and writing software scripts, \agentname completes the same task in about 2 minutes per simulation. At current GPT-4 API rates, a full extraction consumes around \$0.25 per paper. Additionally, \agentname can run on smaller, locally executable language models with no monetary cost and an increased time cost. We report detailed numbers in Table \ref{tab:appendix_time_cost} in Appendix \ref{appendix_time_cost}.

\subsection{Post-Simulation Processing}
The output of simulation software typically consists of particle data, including positions, velocities, masses and optional quantities such as internal energy and star formation rate depending on the physical models enabled. These particles represent matter components in the universe, and the evolution over cosmic time encodes the formation of large-scale structures such as filaments, voids and halos. Some preliminary analysis are crucial for validating and interpreting simulation results:

\begin{itemize}[leftmargin=*]
    \setlength\itemsep{-0.9em}
    \setlength\parskip{1em} 
    \item \textbf{Matter Power Spectrum:} Quantifies the statistical distribution of matter at different scales, sensitive to cosmological parameters such as $\Omega_m$, $\sigma_8$, and $n_s$. Comparing the measured power spectrum with theoretical expectations helps to assess whether the simulation correctly reproduces the output we want.
    \item \textbf{Density Visualization:} Provides intuitive insight into particle distribution, particularly useful for identifying issues like incorrect box sizes or physical model settings. 
    \item \textbf{Specialized Packages:} Generates code for specialized cosmology tools using sample code or minimal user input.
\end{itemize}

The Analysis Code Writer agent automatically provides the user with Python scripts designed to facilitate preliminary analysis of the complex and non-straightforward simulation output. As shown in Figure \ref{fig:post_processing}, the generated code processes the simulation output using various packages to produce the figures described above. Due to the lengthy running time of the simulation software, we were only able to perform visualization analysis and evaluation on a subset of our annotate dataset. The code provided by the Analysis Code Writer agent is highly reliable, with an execution rate of 100\% in the evaluation subset. 

\section{Conclusions}
In this paper, we propose \agentname, a multi-agent system that could accelerate physicist research for cosmological research by automatically performing parameter extraction from user-uploaded paper and simulation setup with preliminary analysis. We demonstrate the system's ability to accurately extract parameters from various simulations and translate them into valid software configuration files. Through benchmark evaluations, \agentname achieves F1 score of 98\%, showing its utility in improving reproducibility, reducing human workload and accelerating the research pipeline. We envision extending \agentname to support additional simulation engines, incorporating more advanced reasoning techniques to interactively assist the researcher during post-simulation analysis. Our system and dataset are released to support further development.

\section*{Limitations}
Due to time constraints, we only annotated one version of the executable files from literature. We are committed to annotating more data points and executable versions in the future. \agentname currently works on a limited number of pre-trained models and simulation software, and we are committed to supporting more models and software variants in the future. 

\newpage

\section*{Ethics Statement}
All models used in our system are commercially available and operated via the OpenAI API under their usage policies. No private, sensitive data were used in this paper. To ensure reproducibility and transparency, we use only publicly available papers, software and user manuals.  


\bibliography{anthology,custom}
\bibliographystyle{acl_natbib}

\clearpage
\appendix
\section{Example script and type of errors}
\label{appendix_example_script}
A correct version of the MP-GADGET simulation script to match the low-resolution simulation in \citealp{ai3}. 
\begin{lstlisting}
"genic": {
      "OutputDir": "./ICs/",
      "FileBase": "LR_100Mpc_64",
      "BoxSize": 100000.0,
      "Ngrid": 64,
      "WhichSpectrum": 2,
      "FileWithInputSpectrum": "./WMAP9_CAMB_matterpower.dat",
      "Omega0": 0.2814,
      "OmegaBaryon": 0.0464,
      "OmegaLambda": 0.7186,
      "HubbleParam": 0.697,
      "ProduceGas": 0,
      "Redshift": 99,
      "Seed": 12345
    }

"gadget": {
      "InitCondFile": "./ICs/LR_100Mpc_64",
      "OutputDir": "./output/",
      "OutputList": "0.333,1.0",
      "TimeLimitCPU": 86400,
      "MetalReturnOn": 0,
      "CoolingOn": 0,
      "SnapshotWithFOF": 0,
      "BlackHoleOn": 0,
      "StarformationOn": 0,
      "WindOn": 0,
      "MassiveNuLinRespOn": 0,
      "DensityIndependentSphOn": 0,
      "Omega0": 0.2814
    }
\end{lstlisting}

An example script with an incorrect simulation box size (Value Error), caused by a mismatch between the units used in the paper and those expected by the simulation software.

\begin{lstlisting}
"genic": {
      "OutputDir": "./ICs/",
      "FileBase": "LR_100Mpc_64",
      <@\hlred{"BoxSize": 100.0,}@>
      "Ngrid": 64,
      "WhichSpectrum": 2,
      "FileWithInputSpectrum": "./WMAP9_CAMB_matterpower.dat",
      "Omega0": 0.2814,
      "OmegaBaryon": 0.0464,
      "OmegaLambda": 0.7186,
      "HubbleParam": 0.697,
      "ProduceGas": 0,
      "Redshift": 99,
      "Seed": 12345
    }
......
\end{lstlisting}

An example script containing an incorrect option that enables gas production in a dark matter only simulation (Type Error), caused by a mismatch between the paper specifications and the generated script.

\begin{lstlisting}
"genic": {
      "OutputDir": "./ICs/",
      "FileBase": "LR_100Mpc_64",
      "BoxSize": 100.0,
      "Ngrid": 64,
      "WhichSpectrum": 2,
      "FileWithInputSpectrum": "./WMAP9_CAMB_matterpower.dat",
      "Omega0": 0.2814,
      "OmegaBaryon": 0.0464,
      "OmegaLambda": 0.7186,
      "HubbleParam": 0.697,
      <@\hlred{"ProduceGas": 1,}@>
      "Redshift": 99,
      "Seed": 12345
    }
    ......

\end{lstlisting}

An example of script containing an incorrect variable name that mismatch with the one in software user manual. (Hallucination)

\begin{lstlisting}
"genic": {
      "OutputDir": "./ICs/",
      "FileBase": "LR_100Mpc_64",
      "BoxSize": 100.0,
      "Ngrid": 64,
      "WhichSpectrum": 2,
      "FileWithInputSpectrum": "./WMAP9_CAMB_matterpower.dat",
      "Omega0": 0.2814,
      "OmegaBaryon": 0.0464,
      "OmegaLambda": 0.7186,
      "HubbleParam": 0.697,
      "ProduceGas": 1,
      "Redshift": 99,
      "Seed": 12345,
      <@\hlred{"FinalRedshift": 0}@>
    }
......
\end{lstlisting}

\section{Evaluation Protocol}
\label{Appendix: Eval}
We define our parameter-level metrics as follows:
\begin{itemize}[leftmargin=*]
\setlength\itemsep{-0.9em}
\setlength\parskip{1em} 
  \item \textbf{True Positives (TP):} Number of extracted parameters whose names and values are exactly correct.
  \item \textbf{False Positives (FP):} Number of extracted parameters with incorrect values/settings .
  \item \textbf{False Negatives (FN):} Number of required parameters that are missing from the extraction output.
\end{itemize}

Our primary evaluation metric is the \textbf{F$_1$ Score}, which captures the overall balance between precision and recall in all extracted parameter instances.
\[
\text{F}_1 = 2 \times \frac{\text{Precision} \times \text{Recall}}{\text{Precision} + \text{Recall}}.
\]

Precision and recall are defined as:
  \[
    \mathrm{Precision} = \frac{\mathrm{TP}}{\mathrm{TP} + \mathrm{FP}}
  \]
  \[
      \mathrm{Recall} = \frac{\mathrm{TP}}{\mathrm{TP} + \mathrm{FN}}.
  \]

A higher F$_1$ indicates more accurate extractions with fewer missing or incorrect parameters.

We categorize error cases into the following types:

\begin{itemize}[leftmargin=*]
\setlength\itemsep{-0.9em}
\setlength\parskip{1em} 
  \item \textbf{Value Error:} The extracted parameter exists but its numerical value is incorrect. This includes errors due to unit mismatch, incorrect scaling, or misinterpretation of scientific notation.
  
  \item \textbf{Type Error:} A parameter is extracted from an incompatible simulation context. (e.g. hydrodynamic settings mistakenly used in a dark matter only simulation)
  
  \item \textbf{Hallucination:} The system outputs parameters that do not appear in the user manual, inventing values or name unsupported by the source.
\end{itemize}

Each of these types of error is reported as the average number of errors per simulation.

\section{Additional Experiments and Results}
\label{Appendix: autoEval}
We provide the automatic evaluation results on \agentname and the baselines in Table \ref{tab:appendix_accuracy} and Table \ref{tab:appendix_error-analysis}. The evaluation results are slightly worse for the baselines compared to the human evaluation, as the automatic evaluation does not consider all possible executable variations of the input file. We provide the automatic evaluation results on \agentname using different backbone models in Table \ref{tab:qwen_accuracy} and Table \ref{tab:qwen_error-analysis}.
\begin{table}[ht]
\centering
\setlength{\tabcolsep}{6pt} 
\begin{tabular}{@{}lccc@{}}
\toprule
\textbf{Method} & \textbf{Micro-F1} & \textbf{Precision} & \textbf{Recall} \\
\midrule
CoT (1-Agent)  & \underline{91.27} & \underline{85.84} & \underline{97.44} \\
EoT (2-Agent)  & 90.69 & 84.94 & 97.27 \\
Ours (2-Agent) & \textbf{98.13} & \textbf{97.77} & \textbf{98.50} \\
\bottomrule
\end{tabular}
\caption{Performance comparison of \agentname with baseline methods on the cosmological simulation dataset. We report Micro-F1 score, Precision, and Recall as percentages. Higher values indicate better performance. The best-performing methods are bolded, and the second-best are underlined.}
\label{tab:appendix_accuracy}
\end{table}

\begin{table}[ht]
\centering
\setlength{\tabcolsep}{4pt} 
\begin{tabular}{@{}lcc@{}}
\toprule
\textbf{Method} & \textbf{Value Error} & \textbf{Type Error}  \\
\midrule
CoT (1-Agent)  & \underline{1.76} & 1.00  \\
EoT (2-Agent)  & 1.97 & \underline{0.95} \\
Ours (2-Agent) & \textbf{0.40} & \textbf{0.05}\\
\bottomrule
\end{tabular}
\caption{Performance comparison of \agentname with baseline methods on the cosmological simulation dataset, in terms of average number of errors made per simulation. Each error type is reported as the average number of errors per simulation. Lower values indicate better performance. The best-performing methods are bolded, and the second-best are underlined.}
\label{tab:appendix_error-analysis}
\end{table}

\begin{table}[ht]\small
\centering
\setlength{\tabcolsep}{6pt} 
\begin{tabular}{@{}lccc@{}}
\toprule
\textbf{Method} & \textbf{Micro-F1} & \textbf{Precision} & \textbf{Recall} \\
\midrule
\agentname & \multirow{2}{*}{98.13} & \multirow{2}{*}{97.77} & \multirow{2}{*}{98.50} \\
(GPT-4) & & & \\
\cmidrule(lr){2-4}
\agentname & \multirow{2}{*}{81.23} & \multirow{2}{*}{70.16} & \multirow{2}{*}{96.10} \\
(Qwen3-4B) & & & \\
\bottomrule
\end{tabular}
\caption{Performance comparison of \agentname using Qwen3-4B as the backbone model and GPT-4 as the backbone model. Experiments are conducted on the cosmological simulation dataset. We report Micro-F1 score, Precision, and Recall as percentages.}
\label{tab:qwen_accuracy}
\end{table}

\begin{table}[ht]\small
\centering
\setlength{\tabcolsep}{4pt} 
\begin{tabular}{@{}lcc@{}}
\toprule
\textbf{Method} & \textbf{Value Error} & \textbf{Type Error}  \\
\midrule
\agentname (GPT-4) & 0.40 & 0.05 \\
\agentname (Qwen3-4B) & 3.05 & 2.59 \\
\bottomrule
\end{tabular}
\caption{Error analysis of \agentname using Qwen3-4B as the backbone model and GPT-4 as the backbone model. Experiments are conducted on the cosmological simulation dataset. Each error type is reported as the average number of errors per simulation.}
\label{tab:qwen_error-analysis}
\end{table}

\begin{table}[ht]\small
\centering
\setlength{\tabcolsep}{4pt} 
\begin{tabular}{@{}lccc@{}}
\toprule
\textbf{Backbone} & \textbf{Average Time} & \textbf{Average Cost} \\
\textbf{Model} & \textbf{(seconds)} & \textbf{(\$)} \\
\midrule
GPT-4  & 124 & \textbf{0.25} \\
Qwen3-4B  & 406 & - \\
\bottomrule
\end{tabular}
\caption{Average time and cost per paper of \agentname using GPT4 and Qwen3-4B, respectively}
\label{tab:appendix_time_cost}
\end{table}

\section{Time and Cost Analysis}
\label{appendix_time_cost}
We provide the average time and cost of \agentname using GPT-4 and Qwen3-4B as the backbone model, respectively. For GPT-4, we do direct API calling; for Qwen3-4B, we run experiments on a single NVIDIA A40 GPU and report the time cost.

\end{document}